\documentclass[doublecol]{epl2}
\usepackage{graphicx}
\usepackage{color}
\usepackage{amssymb}
\usepackage{amsmath} 
\usepackage[T1]{fontenc}
\definecolor{dgreen}{rgb}{0., 0.5, 0.}

\title{Variational approach for interacting ultra-cold atoms in arbitrary one-dimensional confinement}

\author{ Przemys\l{}aw Ko\'{s}cik\inst{1}, Marcin P\l{}odzie\'n\inst{2}, Tomasz Sowi\'nski\inst{2}
}
\shortauthor{P. Ko\'{s}cik, M. P\l{}odzie\'n, T. Sowi\'nski}
\institute{
  \inst{1}Institute of Physics, Jan Kochanowski University, ul.  \'{S}wi\k{e}tokrzyska 15, PL-25406 Kielce, Poland\\
  \inst{2}  Institute of Physics, Polish Academy of Sciences, Aleja Lotnikow 32/46, PL-02668 Warsaw, Poland
}

\abstract{Standard analytical construction of the many-body wave function of interacting particles in one dimension, beyond mean-field theory, is based on the Jastrow approach. The many-body interacting ground state is build up from the ground state of the non-interacting system and the product of solutions of the corresponding interacting two-body problem. However, this is possible only if the center-of-mass motion is decoupled from the mutual interactions. In our work, based on the general constraints given by contact nature of the atom-atom interactions, we present an alternative approach to the standard construction of the \textit{pair-correlation} wave-function. Within the proposed ansatz, we study the many-body properties of trapped bosons as well as fermionic mixtures and we compare these predictions with the exact diagonalization approach in a wide range of particle numbers, interaction strengths, and different trapping potentials.}

\begin{document}

\maketitle


The tremendous progress in experimental techniques has opened up the possibilities to control properties of various artificial quantum systems composed of ultra-cold interacting particles \cite{LewensteinBook}. In particular, quasi-one-dimensional systems of atoms with precisely controlled particle number have been experimentally investigated \cite{Kinoshita2004,Selim2008,Selim2009,Selim2010a,Selim2010b,serwane2011deterministic,zurn2012fermionization}. Such systems allow the experimental verification of the different theoretical predictions. The exactly solvable model of infinitely repulsive bosons, Tonks-Girardeau (TG)  gas, observed recently \cite{Kinoshita2004} allowed to explore the Bose-Fermi mapping \cite{Girardeau1960}. In the regime of finite interactions, the best known analytically solvable model is the celebrated Busch {\it et al.} model \cite{Busch1998} of two indistinguishable particles in a harmonic trap. The generalization of this model to finite-range soft-core interactions has been recently found in  \cite{Koscik2018}. For a larger number of particles, exact solutions are known only for homogeneous systems and they are constructed via the Bethe ansatz \cite{Bethe1931,Guan2013}.
For non-solvable models the construction of the many-body wave function, beyond the mean-field approach, can be done within the interpolatory ansatz \cite{Andersen2016,Pecak2017Ansatz} or in the framework of the \textit{pair-correlated function}, \textit{i.e.}, the approximate variational approach based on the solutions of the corresponding interacting two-body problem \cite{Jastrow1955}. Recently, in the context of ultra-cold atoms, this approach was adopted to bosonic systems with long-range \cite{Cremon2012} and contact interactions \cite{Brouzos2012}  as well as for fermionic mixtures \cite{Brouzos2013TwoCompMixt} in a one-dimensional harmonic trap.
It should be pointed, however, that such a construction is possible only when the solution of the two body-problem can be represented as a product of the center-of-mass and relative motion wave function. Consequently, this conceptually easy approach is very limited.

In this work, we present an alternative construction of the wave function for the many-body ground state based only on a constraint forced by the contact interactions.
Such an approach allows us to consider many-body wave function independently on the two-body solutions.

Let us consider the system of $N$ ultra-cold contact interacting particles in a quasi-one-dimensional trap $V(x)$ described by the Hamiltonian
\begin{equation}\label{Hamiltonian_total}
 {\cal H} = \sum_{i=1}^N\left[-\frac{\hbar^2}{2m}\frac{\partial^2}{\partial x_i^2} + V(x_i)\right] + g\sum_{i<j}\delta(x_i-x_j),
\end{equation}
where the parameter $g$ is an effective strength of contact interactions tunable via Feschbach resonances \cite{Feshbach2010} or strength of the trap in perpendicular directions \cite{Olshanii1998}. Although the Hamiltonian \eqref{Hamiltonian_total} is very general, in the following, we focus on the system of $N$ spinless bosons or on the two-component (pseudo) spin-1/2 mixture of $N=N_\uparrow+N_\downarrow$ fermions. Let us note that the interaction term in Eq. \eqref{Hamiltonian_total} contributes  only when the spatial wave function is symmetric under particle exchange. For fermions with the same spin the wave function is always antisymmetric. Therefore the interaction term is not vanishing only between particles of opposite spins. For further convenience we introduce the position vector $\boldsymbol{r}=(x_1,\ldots,x_N)$.

It is important to note that inter-particle interactions modeled by the Dirac $\delta$ pseudo-potential introduce very specific, non-analytical conditions  to any eigenstate $\Psi(\boldsymbol{r})$ of the Hamiltonian \eqref{Hamiltonian_total}. These conditions read \cite{1963Lieb}:
\begin{multline}\label{con}
\left.\left(\frac{\partial \Psi}{\partial x_i}-\frac{\partial \Psi}{\partial x_j}\right)\right|_{x_{i}=x_{j}^{+}}-\left.\left(\frac{\partial \Psi}{\partial x_i}-\frac{\partial \Psi}{\partial x_j}\right)\right|_{x_{i}=x_{j}^{-}} \\=2g \left.\Psi(\boldsymbol{r})\right|_{x_{i}=x_{j}}.
\end{multline}
The simplest way to determine the structure of the ground state satisfying these conditions is to rewrite its wave function as a normalized product
\begin{equation}\label{Jastrow_ansatz}
 \Psi_G(\boldsymbol{r}) =  {\cal N}\,\Phi(\boldsymbol{r}) \phi(\boldsymbol{r}),
\end{equation}
of the analytical function $\Phi(\boldsymbol{r})$ encoding information about the external confinement and the non-analytical function $\phi(\boldsymbol{r})$ which directly fulfills requirements caused by contact interactions. Representation of the ground-state wave function as a product \eqref{Jastrow_ansatz} is very general and can be realized in many different ways. However, since all mutual interactions considered affects only two selected particles, one can restrict the whole analysis only to situations when the function $\phi(\boldsymbol{r})$ is a {\it product of correlated pairs}
\begin{equation} \label{phitrial}
\phi(\boldsymbol{r}) = \prod_{i<j} \varphi\left( \frac{x_i-x_j}{\sqrt{2}} \right),
\end{equation}
with function $\varphi(x)$ having discontinuous derivative at $x=0$ and satisfying
\begin{equation}
\lim_{\eta\to0^{+}}\left[\varphi'(\eta)-\varphi'(-\eta)\right]=\sqrt{2}g\varphi(0).
\end{equation}
This condition assures that the corresponding total wave function satisfies the conditions \eqref{con}.
The general framework presented above can be utilized for different systems and confinements straightforwardly by selecting functions $\Phi(\boldsymbol{r})$ and $\phi(\boldsymbol{r})$ from some family of variational wave functions satisfying the mentioned conditions.

In the following, we will adopt this procedure in the case of spinless bosons and mixture of fermions showing that there is a huge freedom in choosing appropriate trial functions. The case of harmonically trapped spinless bosons ($V(x)=m\Omega^2 x^2/2$) was deeply analyzed in the literature \cite{Cremon2012,Brouzos2012,Brouzos2013TwoCompMixt}. Following Jastrow, in this case it is typically assumed that the correlated pair function $\varphi(x)$ can be chosen as the ground-state solution of the corresponding two-body problem with respect to the relative motion while analytical part $\Phi(\boldsymbol{r})$ is assumed to be the ground-state wave function of the non-interacting particles.
This approach is doable since in the case of the harmonic potential the ground-state wave function for the relative motion  of the two-body Hamiltonian is known and it can be  expressed in terms of the confluent hypergeometric function \cite{Busch1998} (without loosing generality we set $m=\hbar=\Omega=1$)\begin{equation}\label{Busch0}
\psi(x_1,x_2) =\mathrm{e}^{- (x_1^2+x_2^2)/2}\,\mathrm{U}\left(-{\epsilon\over 2},\frac{1}{2},\frac{(x_1-x_2)^2}{2}\right),
 \end{equation} and  $\epsilon$ is the smallest  root of the following transcendental equation
$
2\sqrt{2}\,\Gamma(1/2-\epsilon/2) = -g\,\Gamma(-\epsilon/2),
$
where $\Gamma(x)$ is the Euler gamma function. Based on this two-body solution one introduces a variational family of solutions \eqref{phitrial} which automatically satisfy conditions \eqref{con}
\begin{equation}\label{Busch}
 \varphi_\alpha(x) = \mathrm{U}\left(-{\epsilon_\alpha\over 2},\frac{1}{2}, \alpha^2 x^2\right),
 \end{equation}
with $\epsilon_\alpha$ being dependent on a variational parameter $\alpha$ and determined by the constrain
\begin{equation}\label{trans}
2\sqrt{2}\cdot\alpha\cdot\Gamma(1/2-\epsilon_\alpha/2) = -g\cdot\Gamma(-\epsilon_\alpha/2).
\end{equation}
Variational wave function of this form reproduces the exact solutions in the two limiting cases, $g=0$ and $g\rightarrow\infty$, for any $N$. Therefore, it was recently used with a great success to study ground-state properties of a few bosons \cite{Brouzos2012} and adopted further for other harmonically confined systems \cite{Brouzos2012,Zinner2016}. Unfortunately, this straightforward adaptation of the very intuitive Jastrow idea automatically brings many disadvantages. The main numerical problem is related to a quite intricate definition of the hypergeometric function $\mathrm{U}$ being a function of the parameter $\epsilon_\alpha$ which is determined by $\alpha$ via transcendental equation \eqref{trans}. All these lead to an arduous adaptation procedure for finding the best approximation of the ground state of the system. In consequence, variational approach for larger number of particles become impossible.
Moreover, the approach completely fails when other confinements are considered since the construction is based on the exact solution of the relative motion for the two-body problem. While separation of the relative motion is possible only for problems described by quadratic Hamiltonians, straightforward generalization to other trapping potentials is not possible.

At this point, it should be emphasized that the proposed variational wave function constructed from the two-body solutions \eqref{Busch} and non-interacting ground-state $\Phi(\boldsymbol{r})$ is not the unique possible realization of the general requirements expressed by constraints \eqref{con}. For example, starting from works by Brouzos and Schmelcher \cite{Brouzos2012,Brouzos2013TwoCompMixt}, it is very common to consider analytical part of the wave function
$\Phi(\boldsymbol{r})$ not as a ground-state of non-interacting particles but as a function
\begin{equation}\label{niemcy}
\Phi_\alpha(\boldsymbol{r}) =\mathrm{e}^{-R^2/\alpha^2}\prod_{i<j}\mathrm{e}^{-\alpha^2(x_i-x_j)^2/4}.
\end{equation}
depending on the same variational parameter $\alpha$ which is used in the non-analytical part $\varphi_\alpha(x)$. Although this interesting approach explicitly incorporates a decoupling of the center-of-mass motion and makes the analytical part sensitive to the variational parameter, in fact, it only increases a numerical complexity. Simply, its accuracy is similar to that with $\Phi(\boldsymbol{r})$ taken as a ground-state of the non-interacting particles with $\alpha=\sqrt{2/N}$.

To overcome all difficulties described above we propose a quite simple pair-correlation function which does not require any knowledge on the two body-solution, \textit{i.e.},
\begin{subequations} \label{KoscikAn}
\begin{equation} \label{KoscikAnA}
\tilde\varphi_\alpha(x) = \frac{1}{\alpha}\left(1 - \lambda \mathrm{e}^{-\alpha|x|}\right).
\end{equation}
with parameter $\lambda$ chosen appropriately to fulfill discontinuity conditions \eqref{con}:
\begin{equation} \label{constr}
\lambda = \frac{g}{\sqrt{2}\alpha + g}.
\end{equation}
\end{subequations}
\begin{figure}
\begin{center}
\includegraphics[width=0.229\textwidth]{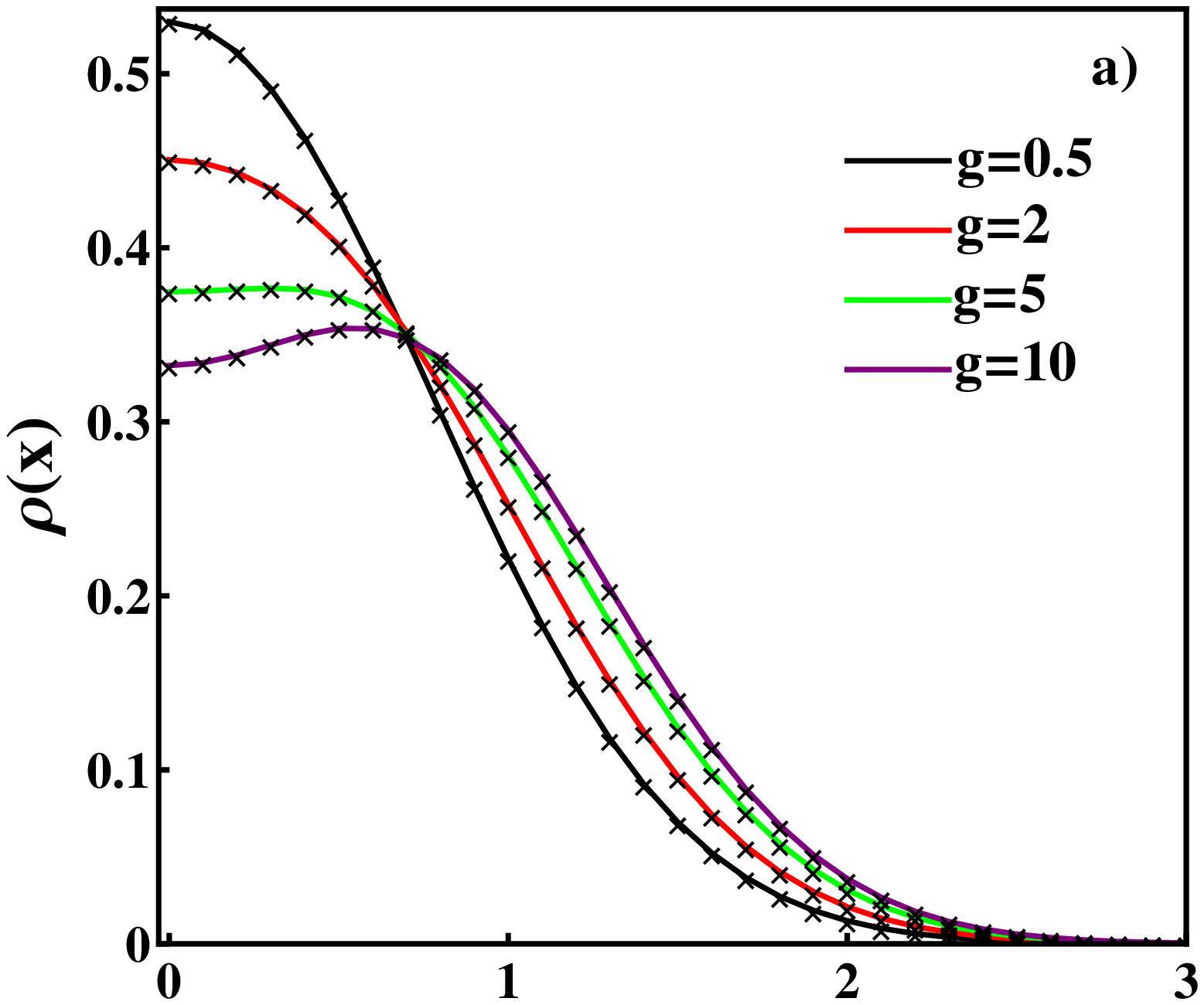}
\includegraphics[width=0.225\textwidth]{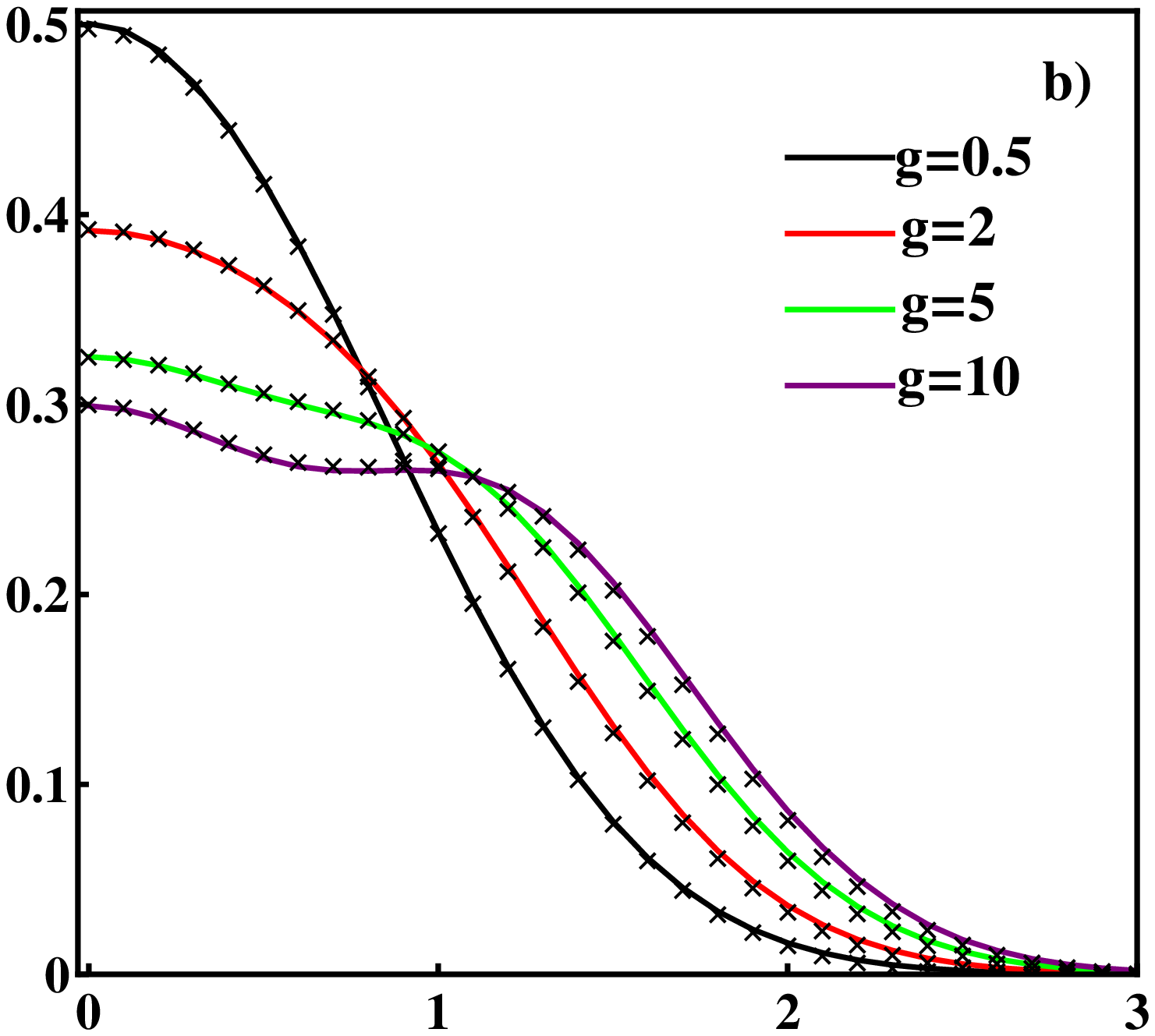}
\includegraphics[width=0.229\textwidth]{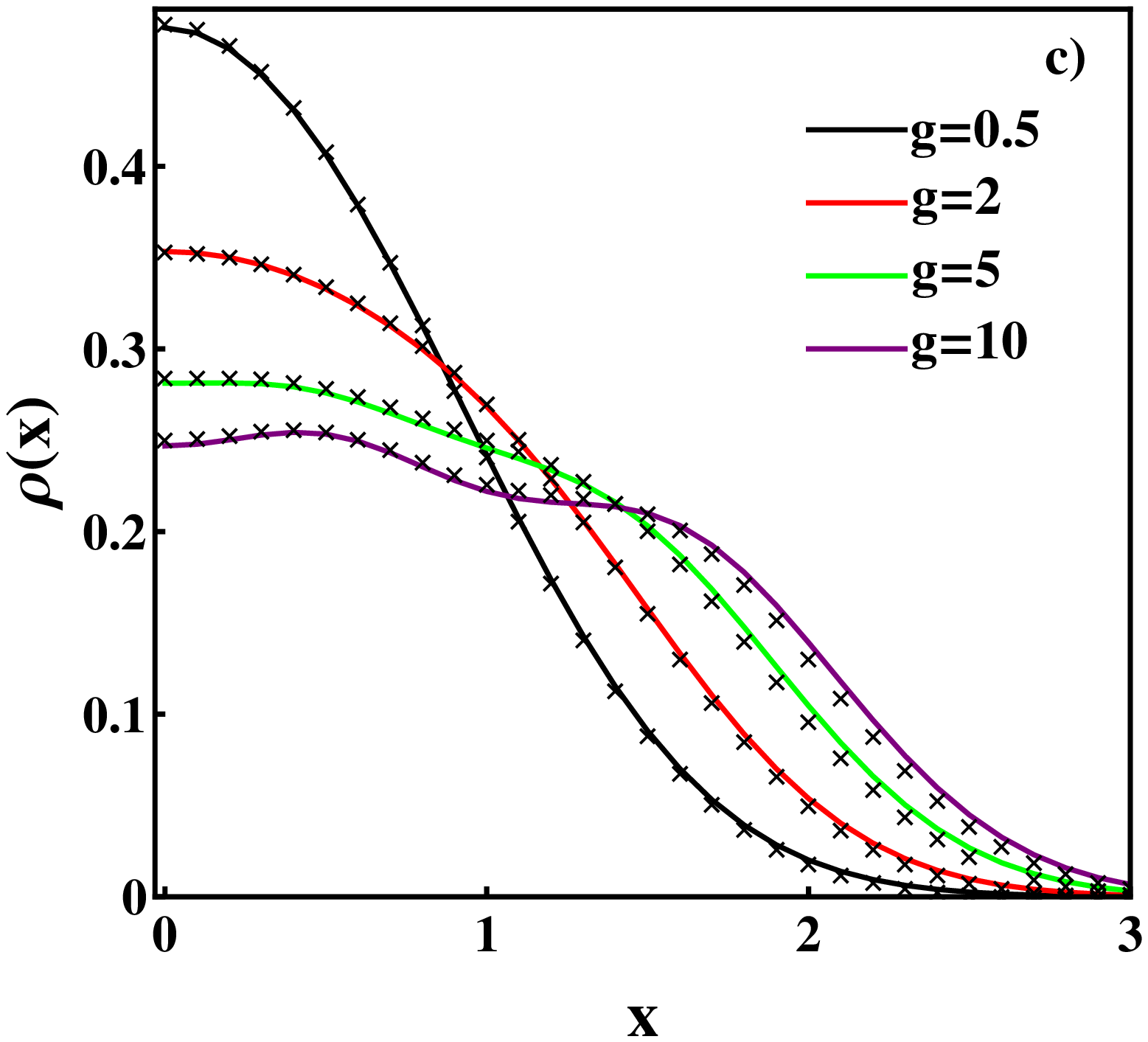}
\includegraphics[width=0.225\textwidth]{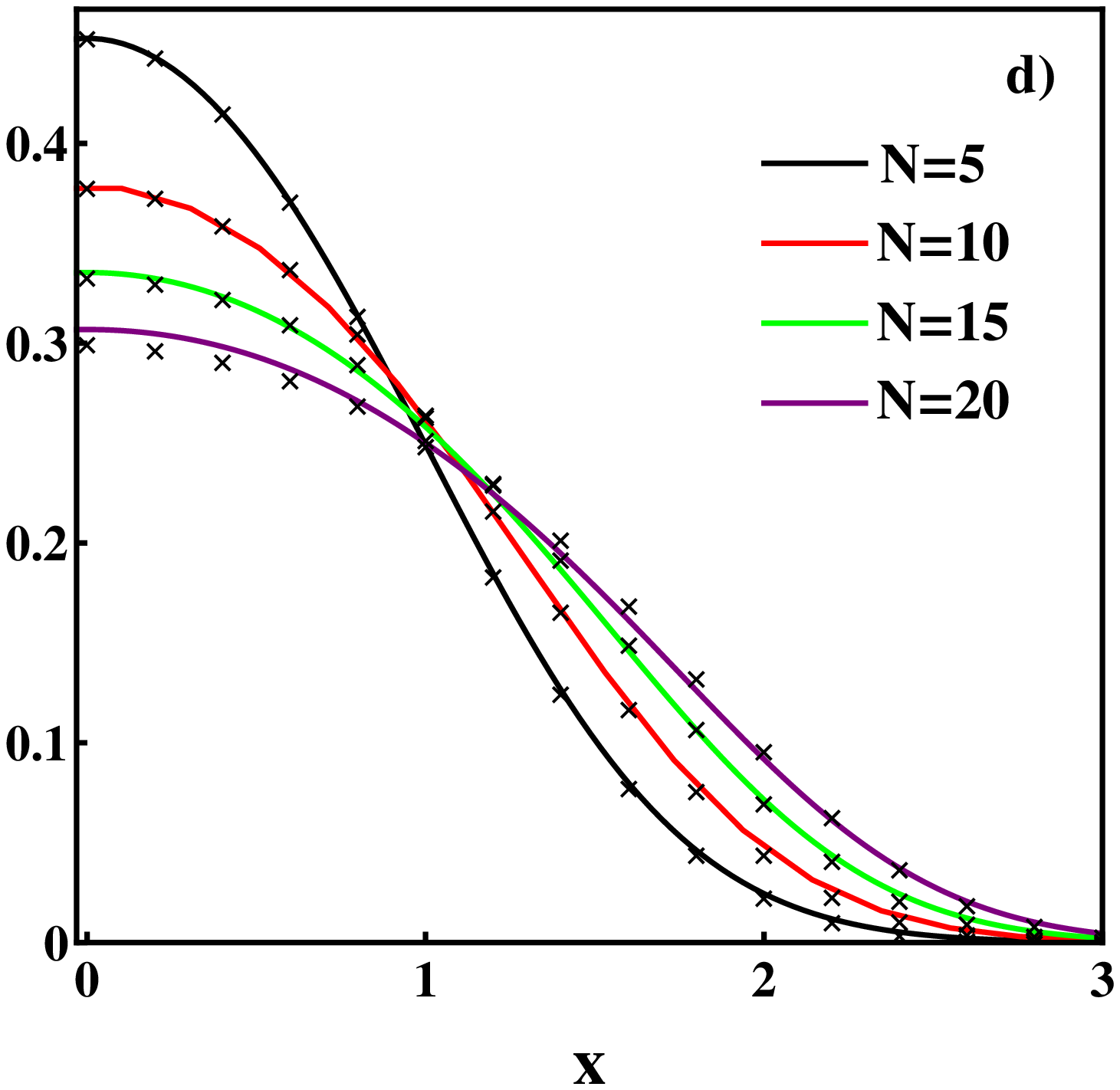}
\end{center}
\caption{\label{Fig1}
Single-particle density profile for interacting bosons confined in a harmonic trap obtained with the exact diagonalization approach (solid lines) and the ansatz (\ref{KoscikAn}) (crossed dots).
Panels $(a)$, $(b)$ and $(c)$ correspond to particle number $N = 2, 3, 4$, respectively. Each panel presents different interaction strength $g = 0.5, 2, 5, 10$ (lines from top to bottom). Panel (d) presents an analogous comparison between  ansatz (\ref{KoscikAn}) and exact diagonalization for fixed interaction strength $g=0.5$ and different particle numbers $N = 5, 10, 15, 20$. Positions and densities are measured in natural units of the harmonic oscillator, $\sqrt{\hbar/m\Omega}$ and $\sqrt{m\Omega/\hbar}$, respectively.}
\end{figure}

It is worth noting that a very similar two-parameter pair-correlation function of the form $\tilde\varphi(x,y)=1-\beta \mathrm{e}^{-\alpha (x^2+y^2)}$ has been successfully used for studying  the  systems of bosons in a two-dimensional isotropic harmonic trap in the presence of a finite-range Gaussian interaction \cite{mutaj}.

Directly by this construction, whenever analytical part $\Phi(\boldsymbol{r})$ is chosen as a non-interacting many-body bosonic ground state, the total many-body wave function \eqref{Jastrow_ansatz} automatically reproduces (up to the normalization factor $\cal N$) the many-body ground-state for vanishing interactions ($g=0$). Moreover, the analytical solution in the TG limit (in a harmonic confinement) is also appropriately captured with this ansatz. In fact, when the system approaches the TG limit ($g\rightarrow\infty$), the variational parameter $\alpha\rightarrow 0$ and consequently $\tilde\varphi_\alpha(x)\rightarrow |x| $. All this means, that the ansatz \eqref{KoscikAn} has all properties of the prior descriptions based on the confluent hypergeometric function. At the same time, it is devoid of numerical difficulties and can be used for a large number of particles with limited numerical resources.

At this point it should be emphasized that the general framework presented above paves the way for generalization to other confinements. This possibility originates in a huge freedom of choosing the form of the pair-correlation function $\phi(\boldsymbol{r})$. Although one can expect that the proposed function \eqref{KoscikAn} is a reasonable choice for any single-well, spatially bounded potential, it certainly fails for other, more complex external confinements. For obvious reasons, there is no general prescription for choosing the best form of the pair-correlation function and one should rely mainly on physical intuition and on numerical studies of trial and error.

The variational trial function \eqref{KoscikAn}, as an approximate description of the many-body ground state, describes appropriately ground-state properties in a whole range of repulsive interactions. We checked that for bosonic systems it recovers the ground-state energy perfectly when compared to the standard approach \cite{Brouzos2012} based on known two-body solutions \eqref{Busch}. It also appropriately describes different single-particle quantities. Indeed, in Fig.~\ref{Fig1} we plot single-particle density profile
\begin{equation} \label{rhoDef}
\rho(x) = \int |\Psi_G(x,x_2,...,x_N)|^2\, \mathrm{d}x_2\cdots\mathrm{d}x_N
\end{equation}
as predicted by the ansatz (crossed dots) for the system of $N$ interacting bosons confined in a harmonic trap. The results are compared with those obtained by a direct exact numerical diagonalization of the many-body Hamiltonian \eqref{Hamiltonian_total} (solid lines). The diagonalization is performed in the lowest energetically Fock basis constructed from single-particle orbitals of a harmonic oscillator \cite{Haugset1998,Plodzien2018}. Additionally, to accelerate the convergence of the method, we apply an optimization strategy based on minimization of the ground-state energy with respect to an effective width of the single-particle basis \cite{Koscik2018opty}. This strategy strongly reduces the number of single-particle orbitals needed to obtain well-converged results.
In panels (a)-(c) in Fig.~\ref{Fig1} we present the results for small number of particles ($N = 2, 3, 4$) and different interactions varying from the perturbative ($g = 0.5$) to the strongly-correlated regime ($g = 10$). In contrast, in panel (d) we present the results for a large number of particles and fixed interaction $g=0.5$. Note that for a larger number of particles the interaction energy is strongly amplified and the system is far from the perturbative regime. Nevertheless, In all cases presented, predictions of the ansatz are almost in a perfect agreement with the exact ones, properly reconstructing the TG limit.


Now let us demonstrate that the ansatz \eqref{KoscikAn} can also be used for other confinements. We focus on the simplest anharmonic trapping potential, {\it i.e.}, the quartic anharmonic oscillator
$V(x) = K x^4$. In this case it is convenient to measure all quantities in natural units of the problem. Namely, if one measures lengths and energies in units of $(\hbar^2/mK)^{1/6}$ and $(K\hbar^4/m^2)^{1/3}$, respectively, then the corresponding many-body Hamiltonian reads:
\begin{equation}
{\cal H}= \sum_{i=1}^N\left(-\frac{1}{2}\frac{\partial^2}{\partial x_i^2} + x_{i}^4\right)+\tilde{g}\sum_{i<j}\delta(x_i-x_j),\end{equation}
where $\tilde g$ is appropriately rescaled strength of interactions in chosen units. Nothe that in this case the corresponding two-body solution is not known and therefore the standard Jastrow approach cannot be adopted. However, as explained before, the adoption of the proposed ansatz is straightforward since the only modification is encoded in the analytical part of the many-body wave function $\Phi(\boldsymbol{r})$. Therefore, to find an optimal many-body ground state, we solve numerically the single-particle problem in selected confinement and we use the lowest energy eigenstate to construct the function $\Phi(\boldsymbol{r})$. Then, we perform the variational optimization of the many-body ansatz \eqref{Jastrow_ansatz} with correlated pair function \eqref{KoscikAn}.
\begin{figure}
\begin{center}
\includegraphics[width=0.239\textwidth]{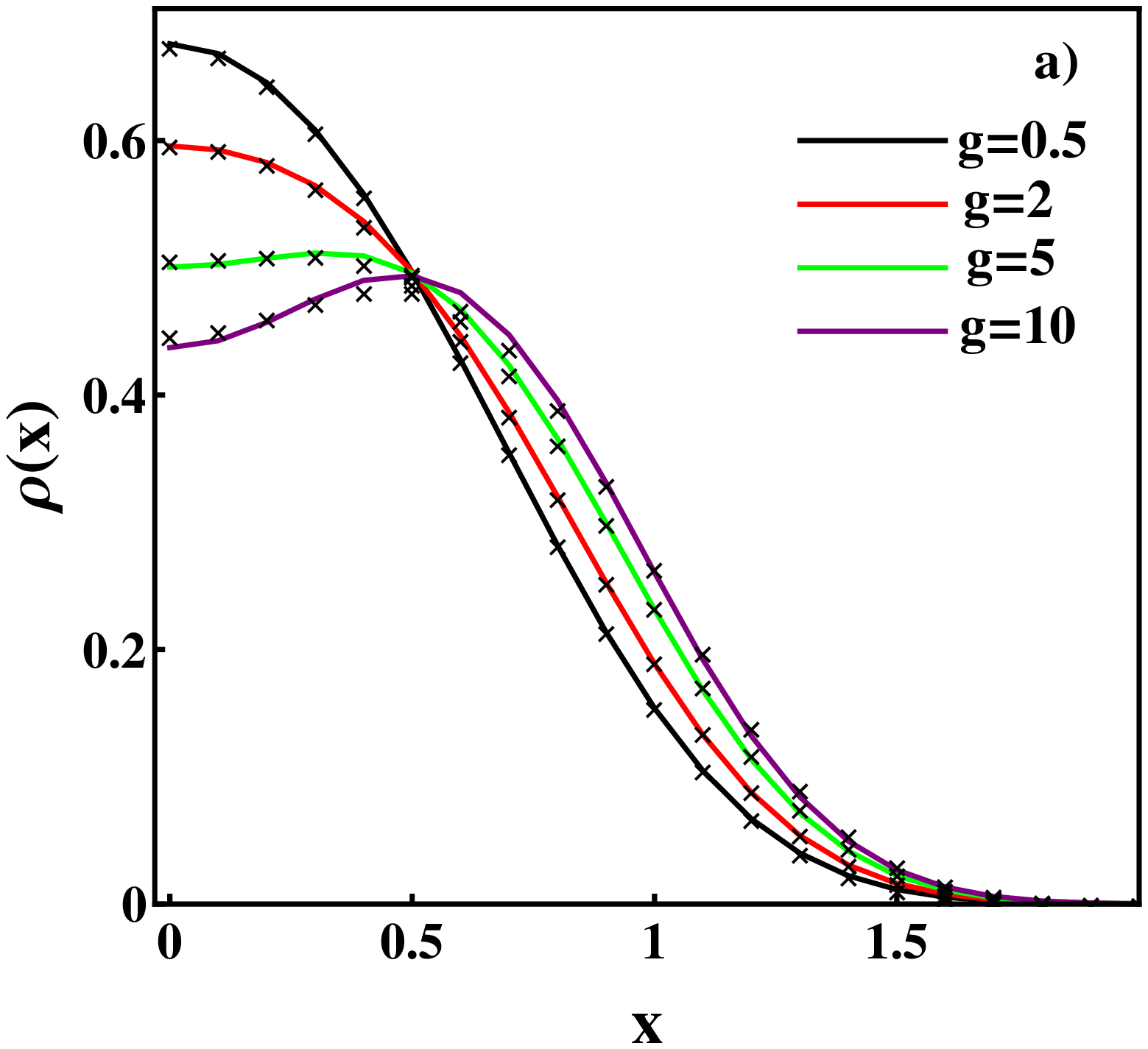}
\includegraphics[width=0.234\textwidth]{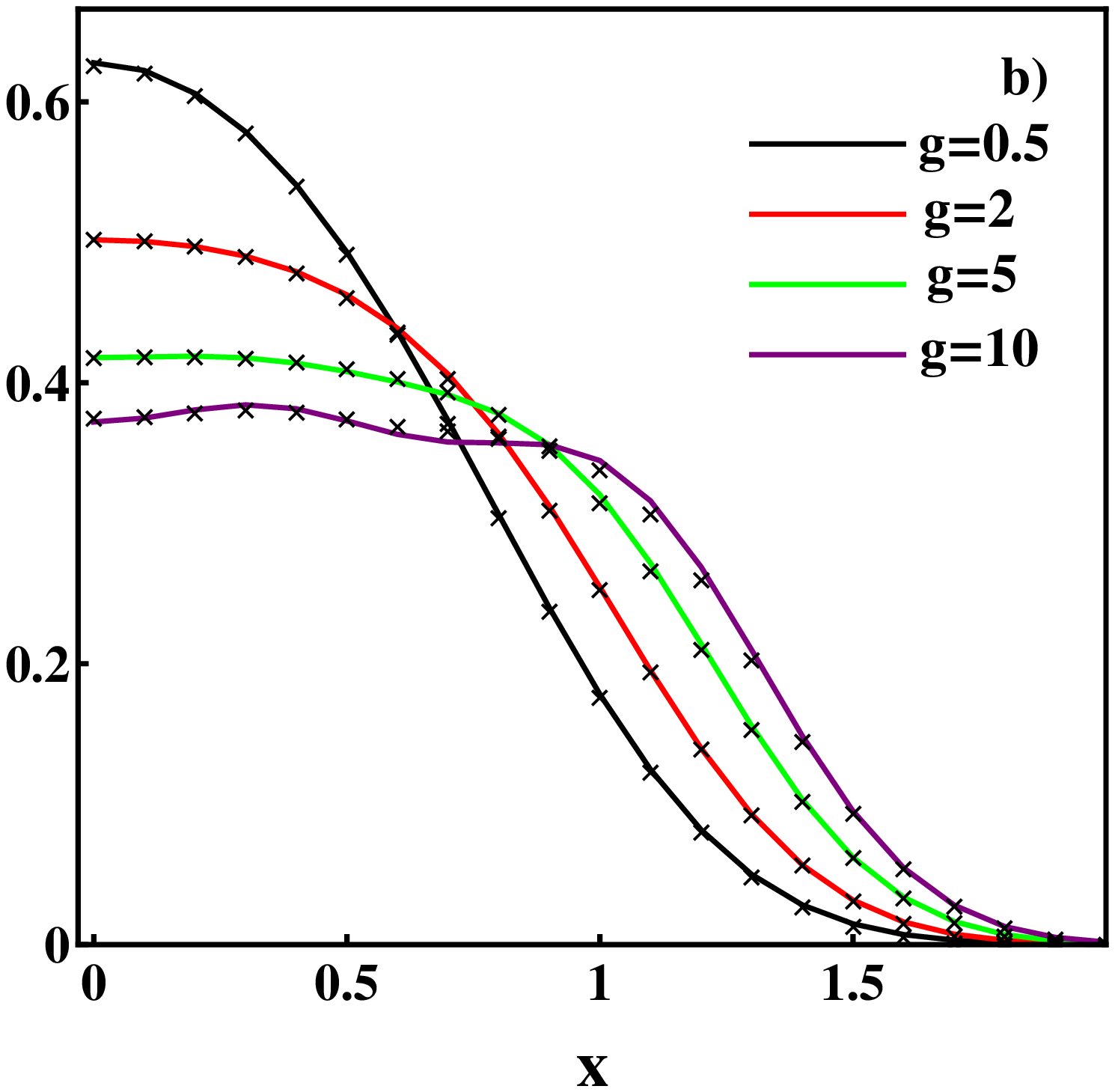}
\end{center}
\caption{\label{Fig2}
Single-particle densities for $N=2$ (left panel) and $N = 4$ (right panel) bosons confined in a potential $V(x) = x^4$ with interaction strengh $g = 0.5, 2, 5, 10$ (lines from top to bottom),
obtained with exact diagonalization (solid lines) and the ansatz \eqref{KoscikAn} (crossed dots). Positions and densities are measured in natural units of the problem, $(\hbar^2/mK)^{1/6}$ and $(mK/\hbar^2)^{1/6}$, respectively.
}
\end{figure}
In Fig.~\ref{Fig2} we present single-particle density profiles of the many-body ground state \eqref{rhoDef} for $N=2$ and $N=4$ particles obtained with the ansatz (crossed dots) and we compare them with those obtained with the optimized exact diagonalization approach described before. As it is seen, the ansatz almost perfectly recovers single-particle shapes for small as well as for strong interactions.

Finally, let us discuss a much more complicated system of $N=N_\uparrow+N_\downarrow$ fermions confined in a harmonic trap. In this case positions of particles form two algebraic vectors $\boldsymbol{r}_\downarrow=(x^\downarrow_1,\ldots,x^\downarrow_{N_\downarrow})$ and $\boldsymbol{r}_\uparrow=(x^\uparrow_1,\ldots,x^\uparrow_{N_\uparrow})$. Since identical fermions cannot occupy the same single-particle orbital, the analytical part of the ground-state wave function $\Phi(\boldsymbol{r})$ is constructed as a product of two Slater determinants of the lowest harmonic oscillator states
$\Phi_\uparrow(\boldsymbol{r}_\uparrow)$ and $\Phi_\downarrow(\boldsymbol{r}_\downarrow)
$, respectively, while the non-analytical part $\phi(\boldsymbol{r})$  is a product of correlated-pair functions for particles belonging to the opposite flavors only.
In the case of the  infinitely strong interactions between opposite spin fermions the ansatz reproduces the exact many-body ground state.
\begin{figure}
\begin{center}
\includegraphics[width=0.239\textwidth]{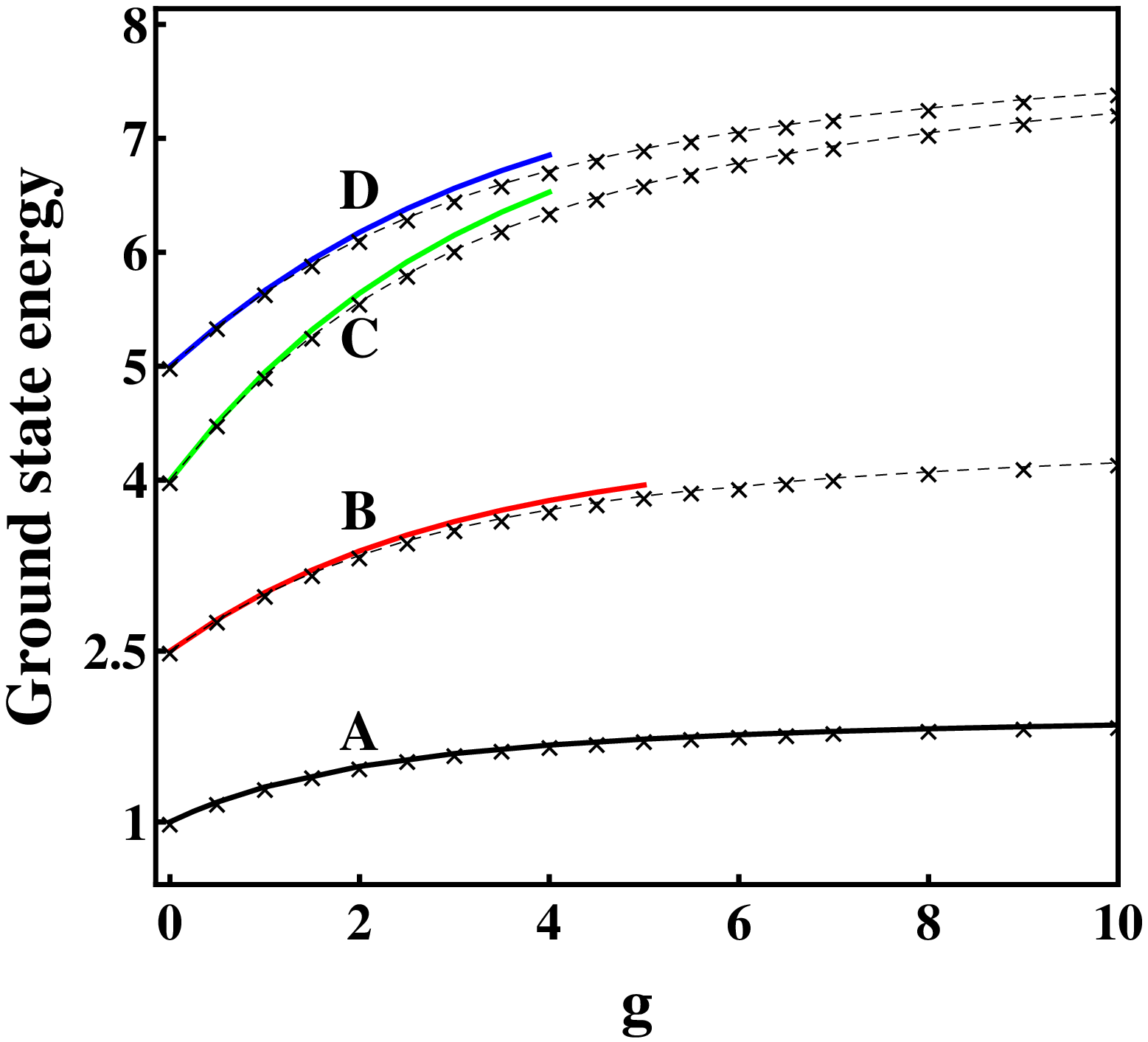}
\includegraphics[width=0.234\textwidth]{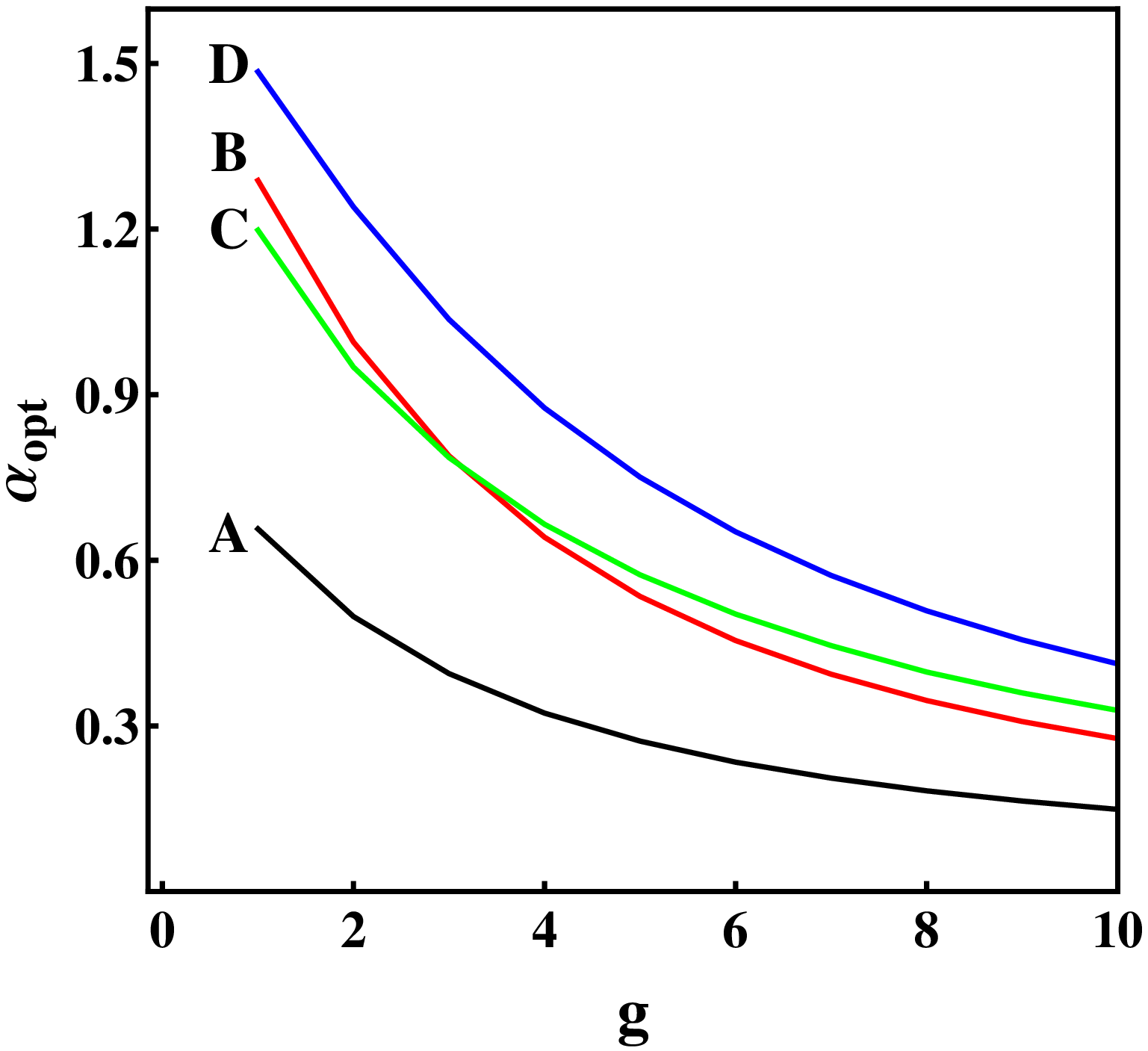}
\textbf{A}: $N_\uparrow=1, N_\downarrow=1$ \\
{\color{red}\textbf{B}}: $N_\uparrow=1, N_\downarrow=2$ \\
{\color{dgreen}\textbf{C}}: $N_\uparrow=2, N_\downarrow=2$ \\
{\color{blue}\textbf{D}}: $N_\uparrow=1, N_\downarrow=3$
\end{center}
\caption{Left panel: Ground-state energy of fermionic mixtures with different number of particles confined in a harmonic trap obtained with the ansatz \eqref{KoscikAn} (crossed dots guided by thin dashed lines). Solid lines corresponds to analytical predictions of the Busch {\it et al.} model \cite{Busch1998} ($N=2$) and  with the exact diagonalization approach ($N>2$). Note that for stronger interactions the exact diagonalization results become overestimated due to the numerical cut-off of a single-particle basis introduced to perform calculations. Right panel: Optimal values of the variational parameter $\alpha$ as a function of interactions for different numbers of fermions confined in a harmonic trap. Interaction strength, energies, and the variational parameter $\alpha$ are measured in natural units of the harmonic oscillator, $(\hbar^3\Omega/m)^{1/2}$, $\hbar\Omega$, and $\sqrt{m\Omega/\hbar}$ respectively.\label{Fig3}
}
\end{figure}
\begin{figure}
\begin{center}
\includegraphics[width=0.239\textwidth]{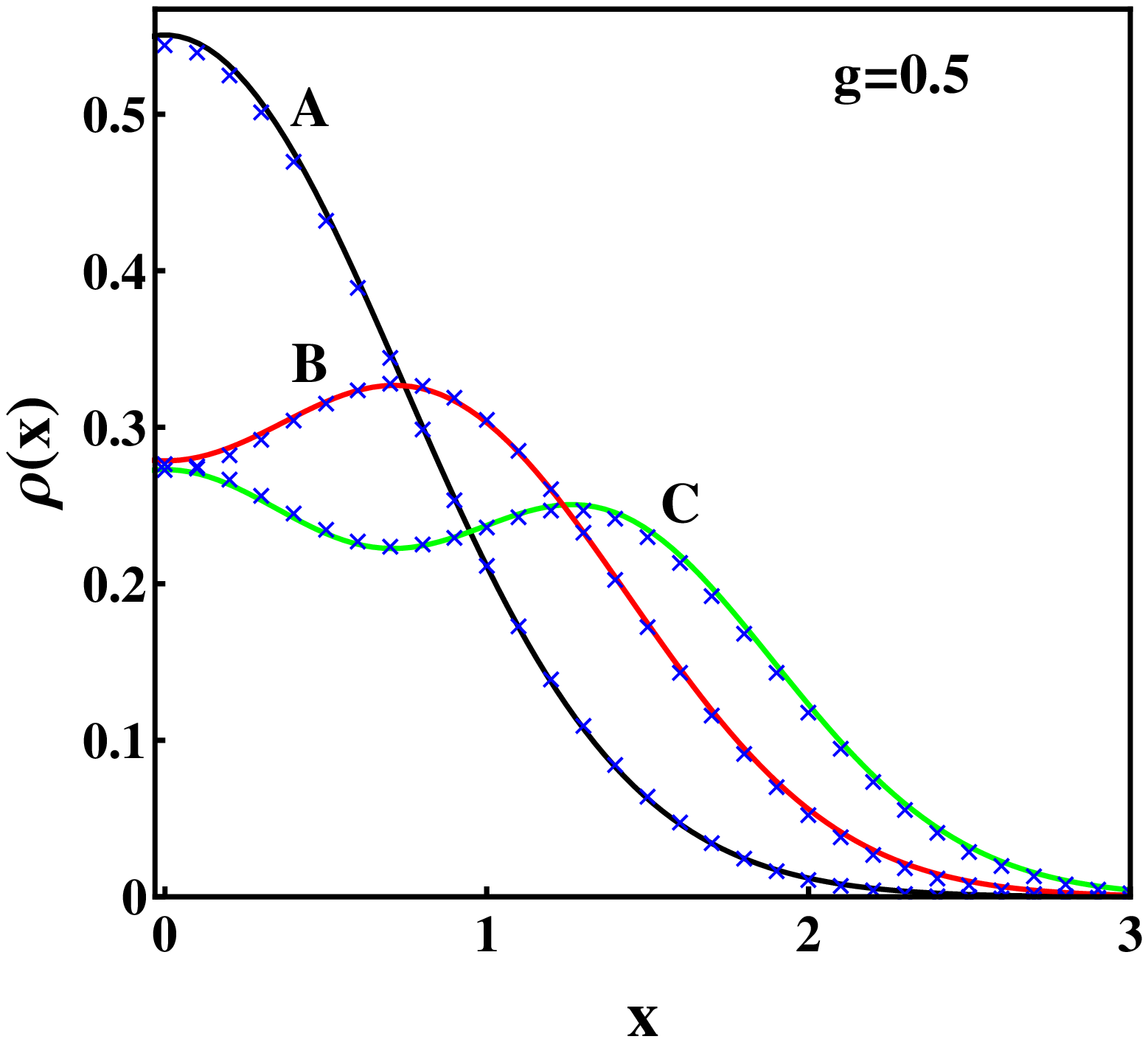}
\includegraphics[width=0.234\textwidth]{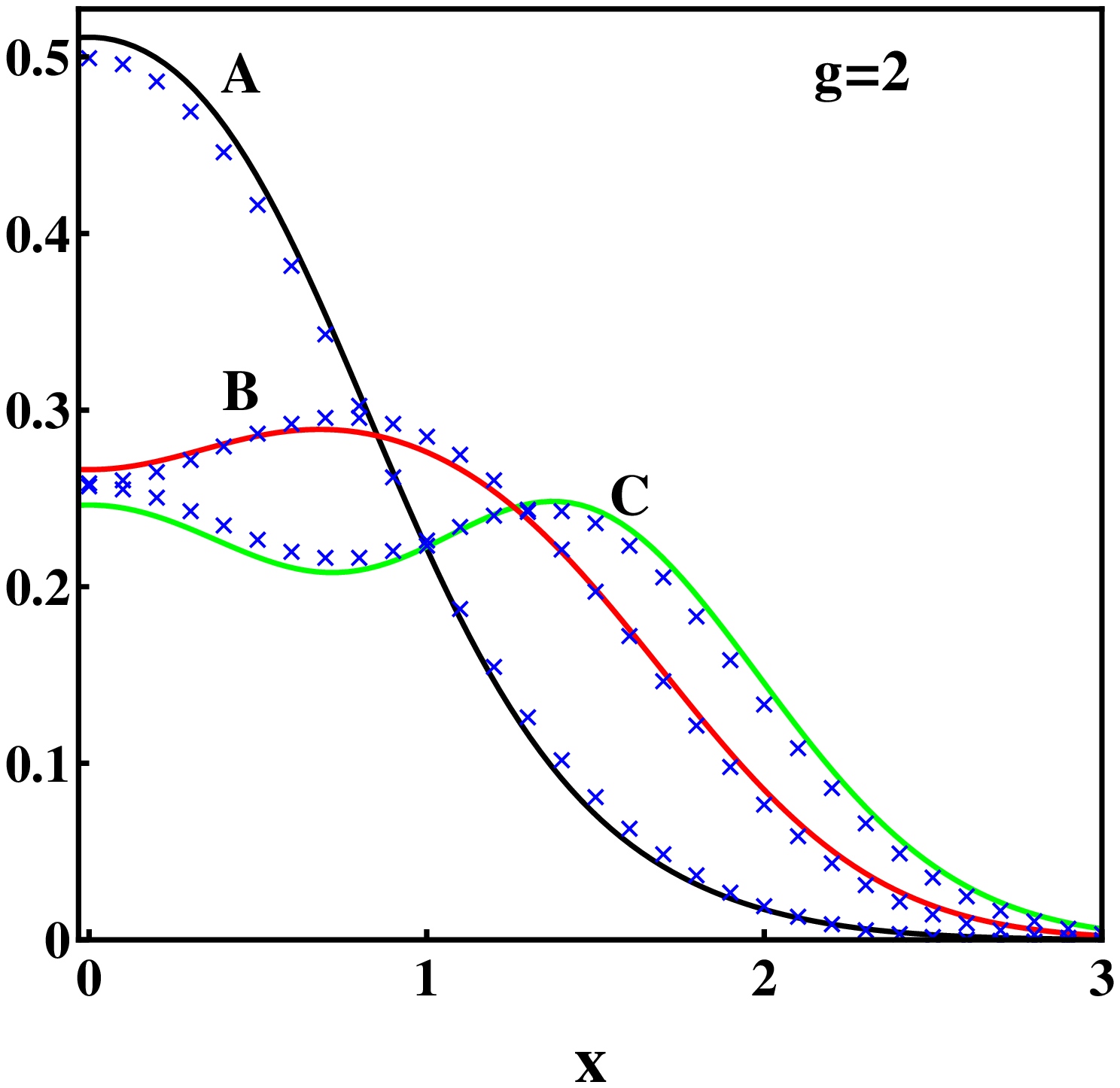}
\end{center}
\caption{\label{Fig4}
Single-particle densities for two-component fermionic mixtures with total particle number $N = N_{\uparrow} + N_{\downarrow} = 4$ for $g = 0.5$ and $g=2$.  Solid lines and crossed dots mark the results obtained from the exact diagonalization and variational ansatz, respectively. The lines A and C mark  the results for the asymmetric mixture    $N_{\uparrow} = 1$, $N_{\downarrow} = 3$,  $\rho_\uparrow(x)$ and $\rho_\downarrow(x)$, whereas   the line B corresponds to the symmetric mixture $N_{\uparrow} = N_{\downarrow} = 2$, $\rho_{\uparrow\downarrow}(x)$. Positions and densities are measured in natural units of the harmonic oscillator, $\sqrt{\hbar/m\Omega}$ and $\sqrt{m\Omega/\hbar}$, respectively. 
}
\end{figure}
For intermediate interactions, the strategy is exactly the same as in the case of bosons, \textit{i.e.}, variational parameter $\alpha$ is found to minimize energy of the system (see Fig.~\ref{Fig3} for details). Having the variational approximation of the ground-state of the system one calculates single-particle density profiles for each component, $\rho_\sigma(\boldsymbol{r}_\sigma)$. In Fig.~\ref{Fig4} we show exemplary results obtained for $N=4$ particles with their different distributions among components ($N_\uparrow-N_\downarrow=0$ or $2$) and different interaction strengths $g$.
Very good agreement of the results with exact diagonalization predictions is clearly visible. Note however, that some deviations are visible for intermediate interactions ($g=2$). They are direct consequences of an immanent inaccuracy of the ansatz which restores the exact solution only in $g=0$ and $g\rightarrow\infty$. Around the center of the trap, where the density of particles is the largest, the density profile is not restored exactly by the ansatz. With increasing interactions, these deviations slowly vanish and the exact result is restored in the TG limit ($g\rightarrow\infty$).

\begin{figure}
\begin{center}

\includegraphics[width=0.239\textwidth]{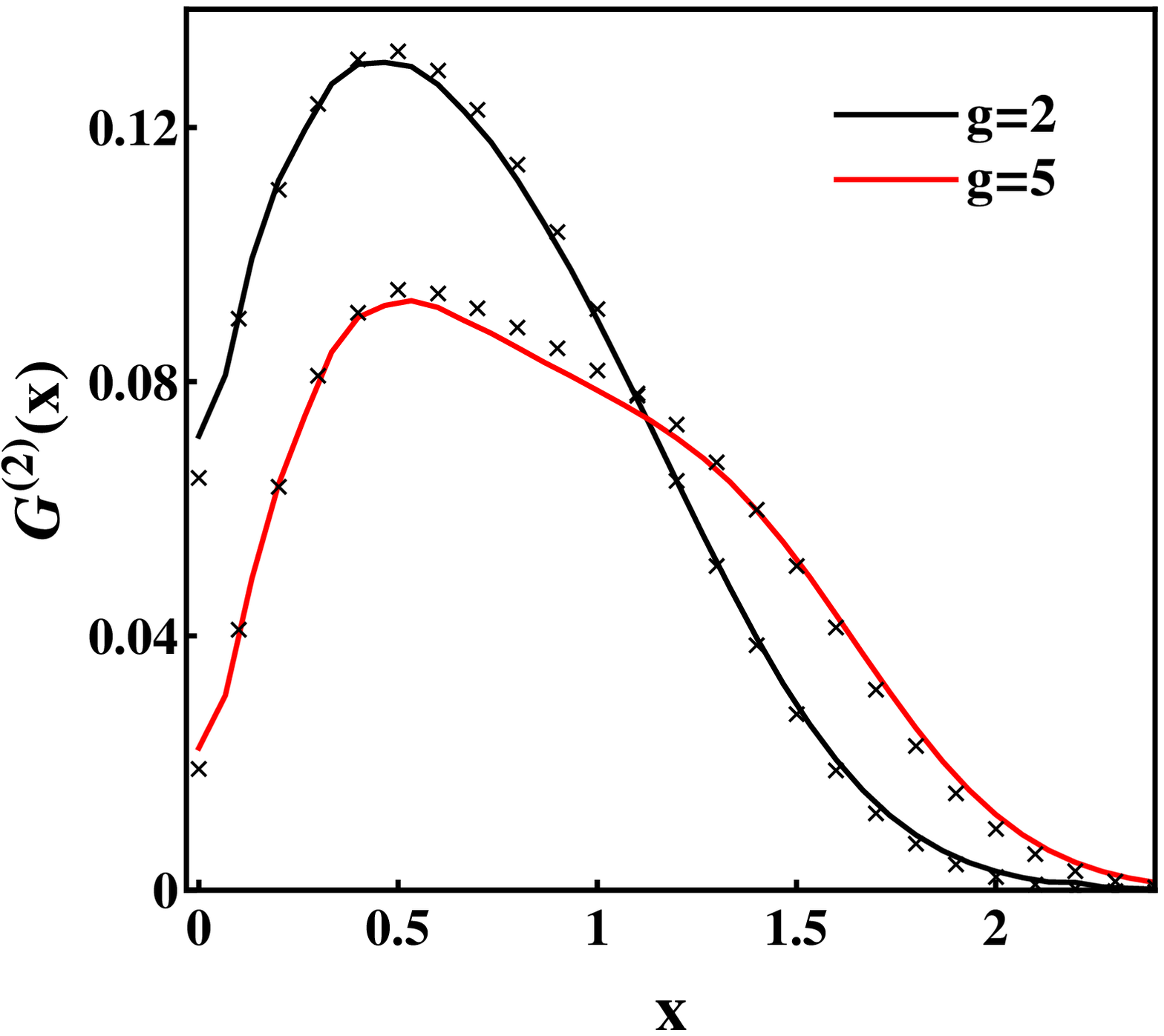}
\includegraphics[width=0.234\textwidth]{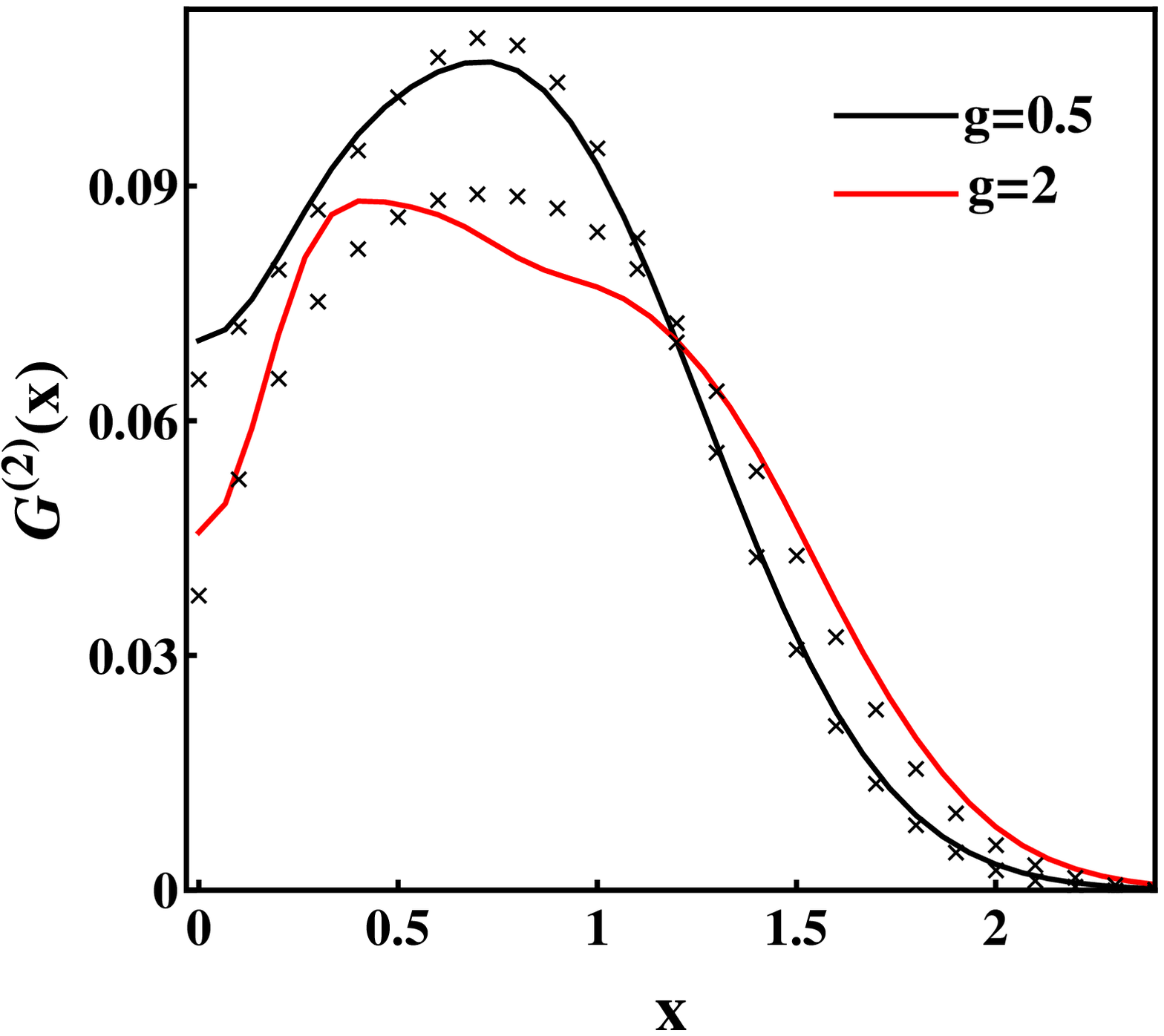}
\end{center}
\caption{\label{Fig5}
Off-diagonal part of the pair-correlation function ${ G}^{(2)}(x)$ for interacting system of $N=4$ bosons (left panel) and $N_\downarrow=N_\uparrow=2$ fermions (right panel) confined in a harmonic trap. Results obtained via the ansatz \eqref{KoscikAn} (crossed dots) are compared with the results obtained with the exact diagonalization of the many-body Hamiltonian (solid lines). Positions and pair-correlation are measured in natural units of the harmonic oscillator, $\sqrt{\hbar/m\Omega}$ and $(m\Omega/\hbar)^{1/4}$, respectively.}
\end{figure}

Finally, let us also show that the ansatz proposed captures not only single-particle quantities but it is also able to predict inter-particle correlations. In the following we focus on correlations encoded in the off-diagonal part of the pair-correlation function
\begin{equation} \label{rhoDef3}
{ G}^{(2)}(x) = \int |\Psi_G(x,-x,...,x_N)|^2\, \mathrm{d}x_3\cdots\mathrm{d}x_N.
\end{equation}
This quantity can be straightforwardly interpreted as a probability density of finding two particles exactly at opposite sides of the trap. Particularly, in Fig. \ref{Fig5} we show ${ G}^{(2)}(x)$ obtained for different four-particle systems (bosons and fermions) confined in a harmonic trap. For fermionic system ($N_{\uparrow} = N_{\downarrow} = 2$) we calculate inter-species correlation by performing integrations in \eqref{rhoDef3} over  $x_{2}^\uparrow$ and $x_{2}^\downarrow$.  It is clearly seen that the results obtained with the ansatz \eqref{KoscikAn} are in good agreement with those obtained by the exact diagonalization of the Hamiltonian. However, some discrepancies are also clearly visible in the case of fermionic mixtures for larger interactions. This is a direct manifestation of the known fact that in the case of fermionic systems inter-particle correlations are much harder to be properly captured by any variational method due to a highly non-trivial role of the quantum statistics \cite{Brouzos2012}.

In the conclusions, we present an alternative approach to the construction of the variational many-body wave function which is based on general restrictions forced by the assumed contact two-body interactions. In contrast to the original Jastrow approach, the proposed ansatz does not require any knowledge about the solution of the corresponding two-body problem. Therefore it can be easily adapted for different trapping potentials and different quantum statistics. It should be pointed out, however, that straightforward generalization to higher dimensions is not possible due to necessary regularization of the contact potential. Predictions of the proposed ansatz are very close to those obtained via the exact diagonalization in a wide range of particle numbers and interaction strengths.

\acknowledgements
This work was supported by the (Polish) National Science Center Grant  No. 2016/22/E/ST2/00555 (MP, TS)

\bibliographystyle{eplbib}

\end{document}